%% file: paper.tex
\documentclass[letterpaper,twocolumn,10pt]{article}
\usepackage[svgnames]{xcolor}
\usepackage{usenix}

\usepackage[scaled=0.8]{beramono} 
\usepackage[normalem]{ulem}
\usepackage{layout}
\usepackage{listings}
\usepackage[font={small},labelfont={bf}]{caption}

\usepackage{booktabs}
\usepackage{graphicx}
\usepackage{subfig}
\usepackage{enumitem}
\usepackage{pifont}
\usepackage{makecell}
\usepackage{titlesec}
\usepackage{siunitx}

\hypersetup{                    
  colorlinks,
  linkcolor={green!20!darkgray},
  citecolor={red!70!black},
  urlcolor={blue!70!black}
}

\newcommand{\keepcomment}{0} 
\newcommand{\code}[1]{\texttt{\detokenize{#1}}}

\makeatletter
\newcommand\thefontsize[1]{{#1 The current font size is: \f@size pt\par}}
\makeatother

\newcommand{\cmark}{\color[HTML]{32CB00} \ding{51}}%
\newcommand{\xmark}{\color[HTML]{FE0000} \ding{55}}%

\newcommand{\tcirc}[1]{\textcircled{\scriptsize \sf #1}}

\makeatletter
\renewcommand{\paragraph}{%
  \@startsection{paragraph}{4}%
  {\z@}{0.1ex \@plus 0.15ex \@minus .1ex}{-0.3em}%
  {\normalfont\normalsize\bfseries}%
}
\makeatother

\titlespacing*{\subsection}
{0pt}{2.5ex plus 0.5ex minus .2ex}{1ex plus .2ex}

\ifnum\keepcomment=1
	\newcommand{\ga}[1]{{\color{red}{[Gianni: #1]}}}
	\newcommand{\rs}[1]{{\color{red}{[Ryan: #1]}}}
	\newcommand{\seb}[1]{{\color{red}{[Sebastiano: #1]}}}
	\newcommand{\as}[1]{{\color{red}{[Alireza: #1]}}}
	\newcommand{\afq}[1]{{\color{red}{[Ashfaq: #1]}}}
	\newcommand{\bs}[1]{{\color{red}{[Brent: #1]}}}
	\newcommand{\delete}[1]{{\leavevmode\sout{#1}}}
\else
	\newcommand{\ga}[1]{\ignorespaces\unskip}
	\newcommand{\rs}[1]{\ignorespaces\unskip}
	\newcommand{\seb}[1]{\ignorespaces\unskip}
	\newcommand{\as}[1]{\ignorespaces\unskip}
	\newcommand{\afq}[1]{\ignorespaces\unskip}
	\newcommand{\bs}[1]{\ignorespaces\unskip}
	\newcommand{\delete}[1]{\ignorespaces\unskip}
\fi

\setlist{nolistsep}

\newcommand{\apigap}{\vspace{0.8eX}}
\makeatletter
\def\@maketitle{%
  \newpage
  \null
  \begin{center}%
    {\Large \bfseries \@title \par}%
    \vskip 1.5ex%
    {\large
      \lineskip .5em%
      \begin{tabular}[t]{c}%
        \@author
      \end{tabular}\par}%
    \vskip 1ex   
  \end{center}%
}
\makeatother

\title{Network-accelerated Active Messages}

\author{
  {\rm Md Ashfaqur Rahaman}\\
  University of Utah\\
  \and
  {\rm Alireza Sanaee}\\
  University of Cambridge\\
  \and
  {\rm Todd Thornley}\\
  University of Utah\\
  \and
  {\rm Sebastiano Miano}\\
  Politecnico di Milano\\
  \and
  {\rm Gianni Antichi}\\
  Politecnico di Milano \&\\Queen Mary University of London\\
  \and
  {\rm Brent E. Stephens}\\
  Google \&\\University of Utah\\
  \and
  {\rm Ryan Stutsman}\\
  University of Utah\\
}

\begin{document}
\maketitle

\input{tex/abstract}
\input{tex/introduction}

\input{tex/background2}

\input{tex/design}
\input{tex/evaluation}

\input{tex/conclusion}

\bibliographystyle{plain}
\bibliography{references}

\end{document}

%% file: tex/abstract.tex
\begin{abstract}

    Remote Direct Memory Access (RDMA) improves
    host networking performance by eliminating software and server CPU involvement.
    However, RDMA has a limited set of operations, is difficult to program, and
    often requires multiple round trips to perform simple application operations (e.g.,\ hash table lookup).
    Programmable SmartNICs provide a different means to offload work from host CPUs to a NIC while supporting application logic. This leaves applications with the complex choice of embedding logic as RPC handlers at servers, using RDMA's limited interface to access server structures via client-side logic, or running some logic on SmartNICs.
    The best choice varies between workloads and over time.

    To solve this dilemma, we present NAAM, {\sl network-accelerated active messages}. NAAM applications specify small, portable eBPF functions associated with messages. 
    Each message specifies what data it accesses using an RDMA-like interface. NAAM runs at various places in the network, including at clients, on server-attached SmartNICs, and server host CPU cores. Due to eBPF's portability, the code associated with a message can be run at any location. Hence, the NAAM runtime can dynamically steer any message to execute its associated logic wherever it makes the most sense based on the data it accesses or current network or CPU load.

To demonstrate NAAM's flexibility, we built several applications, including the MICA hash table and lookups from a Cell-style B-tree. With an NVIDIA BlueField-2 SmartNIC and integrating its NIC-embedded switch, NAAM can run any of these operations on client, server, and NIC cores, shifting load in tens of milliseconds on server compute congestion. NAAM dynamically offloads up to 1.8~million MICA ops/s for YCSB-B and 750,000 Cell lookups/s from server CPUs. Finally, whereas iPipe, the state-of-the-art SmartNIC offload framework, only scales to 8 application offloads on BlueField-2, NAAM scales to hundreds of application offloads with minimal impact on tail latency due to eBPF's low overhead.



\end{abstract}

%% file: tex/introduction.tex
\section{Introduction}
\label{sec:into}

Remote Direct Memory Access (RDMA) can significantly improve the performance of end-host networking
by bypassing software networking stacks and offloading server CPUs~\cite{farm, guo2016rdma, kalia2014using, kalia2019datacenter, zhu2015congestion}. This can dramatically reduce tail latency and CPU usage by avoiding packet processing costs, which consume a large fraction of CPU cycles~\cite{tas,flextoe}.

However, RDMA verbs are notoriously difficult to program, and multiple round trips are typically required for complex operations on server structures (e.g., looking up a key-value pair in a hash table) or they require careful RDMA-specific re-engineering of data structures~\cite{kalia2014using, farm}.
Historically, remote procedure call (RPC) has been the solution to this problem since it runs application-defined logic at servers in response to network messages; however, this comes at the cost of server CPU load.
Recently, SmartNICs have been employed to help solve this by allowing custom operations to be defined and run on NICs~\cite{floem,ipipe,AmaroRMC2020}, generally at server SmartNIC CPUs.


The result is that developers are faced with the complex choice of using RDMA and client-side logic, using RPC and server-side logic, or using custom SmartNIC offloads.
Each model has different trade-offs and bottlenecks, and the best configuration can vary between workloads and even over time.  For example, during normal operation, it is often faster to use a server SmartNIC core or CPU core to walk a remote data structure like a hash table or a B-tree to avoid multiple network round trips.  However, if the remote server cores are overloaded, then it can be better to run the logic at the client.


Building applications that can monitor network and server load and dynamically move logic to optimize performance is difficult.  
To solve this, we propose NAAM, network-accelerated active messages, a new framework for remote memory access and function offloading.
NAAM avoids the need to re-implement the same logic for multiple different platforms because NAAM replaces RDMA with active messages that execute small, portable eBPF functions~\cite{linux-ebpf,bpf1993}.
Instead of writing multiple functions with different RDMA verbs for the same logic (e.g., using both one-sided and two-sided logic), applications can specify a single RPC-like function that can be associated with network messages. This function can access state stored on servers, and it can run custom code; NAAM decides how best to schedule and run these messages' functions on the available compute in the network.


NAAM runs as a software switch at various places in the network, including at clients, on server-attached SmartNICs, and on server host CPU cores. eBPF's portability ensures that each message's code can run at any of these locations, so
the NAAM runtime can dynamically steer any message to execute its associated logic wherever it makes the most sense (e.g., based on load).
In addition to being able to dynamically and incrementally scale workloads to span both SmartNIC and host CPU cores, NAAM can also route around transitory CPU contention at sub-second timescales to improve tail latency.

Although some prior SmartNIC programming frameworks have helped address the problems associated with mapping some work to SmartNIC resources while mapping other work to server host CPUs, they all have important limitations.
For example, Floem pipelines are mapped to SmartNIC and CPU resources, and they do not adapt the placement of work when workloads change~\cite{floem}, creating bottlenecks. iPipe uses an actor-based model to dynamically move work between SmartNIC and host CPU resources within a server~\cite{ipipe}. However, iPipe depends on a special ``on-path'' SmartNIC that integrates compute elements directly into its packet processing pipeline and exposes an exotic software-managed TLB.
Without this special hardware, iPipe's approach breaks down on commodity ``off-path'' SmartNICs like the NVIDIA BlueField-2 that we use, where SmartNIC cores must access packets over a bus (PCIExpress). When running just nine actors on BlueField-2, iPipe's tail response latency increases by three orders of magnitude, and its efficiency breaks down. This is a significant limitation in a multi-tenant data center, where numerous applications are co-located.
NAAM's eBPF runtime is small and has a low cost to enter/exit without specialized hardware, so it efficiently scales to hundreds of inter-isolated user-defined operations with tail response times of tens-of-microseconds, even with popular off-path SmartNICs.

Overall, we show that by reactively shifting active message processing to the ARM cores of a BlueField-2 SmartNIC, NAAM can reduce tail latency by 400$\times$ when host x86 CPUs are contended.  Furthermore, when the NIC ARM cores become overloaded, NAAM automatically reconfigures the integrated BlueField-2 hardware switch to allow the host x86 CPUs to help out, avoiding the bottleneck due to CPU saturation when under heavy load.  We show in the evaluation that this allows NAAM to transparently scale request processing for a disaggregated MICA-based~\cite{lim2014mica} in-memory hash table from the ARM CPUs’ limit of 1.8~Mop/s to 6~Mop/s with assistance from host CPU cores. Finally, as a full application benchmark, we evaluate the lookup path of a re-implementation of the Cell B-tree~\cite{mitchell2013using}. At 1~million requests/s, SmartNIC cores using NAAM achieve comparable performance to using client-driven RDMAs while moving $4.3\times$ less data.
In sum, this paper
\begin{enumerate}[leftmargin=3ex]
    \item describes NAAM's design for a unified interface for application-defined remote memory operations, and its active message-based approach to executing them, which can transparently use heterogeneous platforms that span clients, servers, and SmartNICs;
    \item details NAAM's eBPF runtime, which includes a new scheme that cooperatively suspends and resumes the execution of eBPF functions while still supporting a fast, JIT-based implementation on x86-64 and ARMv8;
    \item describes how this cooperative yield leads to an implementation of NAAM as an efficient software switch that ensures remote state access never introduces blocking within the runtime, which would hurt throughput;
    \item shows that NAAM balances load across host and SmartNIC resources via NIC hardware features while redistributing load when workloads change or interfere at sub-second timescales, resulting in improved tail latency.
\end{enumerate}

%% file: tex/background2.tex
\section{Background \& Motivation}
\label{sec:motivation}

\noindent\textbf{RDMA and its limitations.}
Recently, data center NIC bandwidth has exploded to 400~Gbps, with 800~Gbps coming soon. At these speeds, software-based TCP/IP packet processing costs lead to intolerable overheads. This has led to pervasive use of hardware offloads, including RDMA, which offers direct access to the memory of a remote machine without involving its CPU. RDMA delivers lower latency, lower CPU costs, and improved throughput compared to traditional networking stacks~\cite{farm, dragojevic2015no, mittal2018revisiting}, making it a popular choice for distributed applications including machine learning~\cite{tian2021accelerating, xue2019fast}, distributed transactions~\cite{dragojevic2015no, lu2016high, wei2018deconstructing}, replicated state machines~\cite{poke2015dare}, graph processing~\cite{WANG2023144, shi2016fast, gu2017efficient, freeflow}, and key-value stores~\cite{kalia2014using, mitchell2013using, farm, design-guildlines}.

Despite its advantages, RDMA has limitations that prevent it from being widely adopted. RDMA provides two kinds of interfaces: {\sl one-sided} and {\sl two-sided}.
{\sl One-sided} RDMA operations are handled entirely by the receiver's NIC, eliminating all server CPU involvement, but these operations are mostly limited to reading or writing one contiguous remote memory location.
Operations over more complex structures (e.g.\ a hash table) must be composed of multiple one-sided operations, complicating client logic, making synchronization challenging, and cutting into the benefits of off-loading operations to the NIC.
{\sl Two-sided} RDMA operations still require the receiver's CPU to handle each incoming message. Two-sided operations effectively provide fast RPC, preserving a classic client-server design; however, many applications use small messages at a high rate, so this approach can often result in a bottleneck at the server's CPU.

%
This dichotomy presents a complex decision for application developers, who must consider a range of factors (compute capacity, data access patterns, and the nature of data-dependent operations) to determine the most suitable RDMA model for their specific application. This problem is compounded by the fact that workloads change over time.

\vspace{0.05in}
\noindent\textbf{Enhancing capabilities with SmartNICs.}
\rs{Re commented block here about RDMA cc; do we want to keep these cites somehow?}
Recently, SmartNICs have emerged as an approach to enable more general offloads to NICs.
There are two main classes of compute elements on SmartNICs: FPGAs~\cite{xilinx_40_100g_ethernet_2023,xilinx_virtex_7_fpga_2023}
or SoC cores~\cite{cavium_intelligent_network_adapters_2023, mellanox_bluefield_smart_nic, netronome_nfp6xxx_2023, mellanox_tile_gx_ug}.
SmartNICs also include ASIC accelerators for specific tasks like TCP/UDP offload and encryption/decryption. 
SoC-based SmartNICs generally include capable but simple multicore ARM or MIPS processors.

This categorization further breaks down into two types based on their interaction with network traffic: {\sl on-path} and {\sl off-path}. 
On-path SmartNICs incorporate processing elements directly into the data path, processing every packet through SmartNIC cores. 
Off-path SmartNICs generally include \emph{(i)} a hardware switch that separates compute elements from the data path, allowing for flexible packet routing to either the host system or the SmartNIC cores, \emph{(ii)} a general-purpose multicore CPU capable of running custom applications, and \emph{(iii)} an efficient mechanism for NIC CPUs to access host memory either via DMA or RDMA.

Unlike on-path SmartNICs, off-path variants run full-fledged operating systems, such as Linux, and support kernel-bypass systems like DPDK. They also have substantial onboard NIC memory and are free from the static resource constraints of FPGA-based SmartNICs that can limit the number of concurrently offloadable applications.
Off-path SmartNICs also avoid the time-consuming and manually-intensive process of writing highly NIC-specific, resource-constrained packet processing code. For example, they avoid the need for FPGA synthesis and reconfiguration (including partial reconfiguration)~\cite{vipin2018fpga}.
Considering these strengths, this paper demonstrates an approach for programming offloads that works well for off-path SmartNICs while still admitting an efficient implementation for on-path SmartNICs.


\begin{table}[t]
\centering
\resizebox{\columnwidth}{!}{%
\begin{tabular}{c|cccc}
\toprule
\textbf{System} &
  \begin{tabular}[c]{@{}c@{}}\textbf{Programmability}\end{tabular} &
  \begin{tabular}[c]{@{}c@{}}\textbf{Multi-Tenant}\\ \textbf{Scaling}\end{tabular} &
  \begin{tabular}[c]{@{}c@{}}\textbf{Dynamic}\\ \textbf{Offloading}\end{tabular} &
  \begin{tabular}[c]{@{}c@{}}\textbf{Commodity}\\ \textbf{SmartNIC}\end{tabular} \\ \midrule
    RMC~\cite{AmaroRMC2020}		        & \cmark & \cmark  & \xmark & \cmark      	\\
    PRISM~\cite{BrukePrism2021} 		& \xmark & \cmark 	  & \xmark & \cmark      	\\
    StRoM~\cite{SidlerStorm2020} 		& \cmark		 & \xmark	 	  & \xmark	   & \xmark    		\\
    1RMA~\cite{Singhvi1RMA2020} 		& \xmark & 	\cmark 	  & \xmark	   & \xmark	    	    \\
    iPipe~\cite{ipipe} 					& \cmark & \xmark & \cmark & \cmark  	 \\
    Bertha~\cite{bertha} 				& \cmark & \xmark & \cmark & \xmark	        \\ \midrule
    FaRM~\cite{farm,dragojevic2015no}   & \cmark & \xmark & \xmark    & N/A     		\\
    Snap~\cite{snap} 					& \cmark & \cmark & \xmark    & N/A     		\\
    Splinter~\cite{splinter,asfp,kayak} & \cmark & \cmark & \cmark & N/A \\ \midrule
    \textbf{NAAM}                   	& \cmark & \cmark & \cmark & \cmark  \\ 
\bottomrule
\end{tabular}
}
\caption{A comparison of existing works on custom RDMA extensions and SmartNIC offload frameworks.}

\label{table:comparison}
\end{table}

\subsection{Toward an Active Messaging Framework}
The evolution of RDMA and the advent of SmartNICs present an opportunity to redefine how distributed applications are developed and executed. 
An ideal framework for custom remote memory operations should solve several key challenges. Building a framework that is universally applicable and efficient requires simultaneously addressing all of these challenges.
Here, we outline these challenges; Table~\ref{table:comparison} compares NAAM to other significant contributions in this field.

\paragraph{Programmability.}
Foremost, a remote memory programming framework must
provide an easy-to-use programming model that shifts developers' focus from the complexities of RDMA operations to the logic of their applications. 
Writing RDMA code requires changing every memory access to issue a remote operation including handling errors and synchronization. Synchronization can be especially difficult because ongoing operations on server host CPUs are unaware of RDMA accesses, and additional round trips to set/clear synchronization flags over RDMA are prohibitively expensive.
For example, if a client issues an RDMA read of a value on a server, the returned value may not be atomic due to ongoing server CPU operations that are updating the value (or concurrent RDMA writes).
The resulting code is complex and hard to debug, especially compared to accessing server data structures via RPC, which uses local memory accesses.

PRISM~\cite{BrukePrism2021} and 1RMA~\cite{Singhvi1RMA2020}
introduce new RDMA verbs for more complex operations; however, their set of verbs is still fixed, so their performance benefits don't generalize to all applications. Applications also must still handle errors and synchronization just as with RDMA. FaRM allows applications to mix RPC and RDMA code but it doesn't support SmartNICs or handle multiple tenants~\cite{farm,dragojevic2015no}.

\paragraph{Dynamic Offloading and Resource Management.}
The optimal choice between one-sided and two-sided operations often hinges on the system load~\cite{le2019impact}, necessitating multiple code versions to accommodate varying system conditions~\cite{zhu2015congestion, du2021fast}.
iPipe~\cite{ipipe} uses distributed actors which can be moved between host CPU cores and a SmartNIC to solve this, but it only works with a limited number of actors for off-path SmartNICs.
Bertha~\cite{bertha} lets applications express chains of offloads while choosing their implementations at runtime, but it is restricted to a set of pre-existing kernels and offloads.
RMC~\cite{AmaroRMC2020} lets developers use arbitrary code to specify application-specific offloads to SmartNICs, but it doesn't support online reconfiguration of where those offloads run.
Similarly,  StRoM~~\cite{SidlerStorm2020} allows the synthesis of C++ onto an FPGA-enabled NIC to enable new RDMA verbs, but it doesn't support online reconfiguration to divide work across host and SmartNIC resources.

An ideal framework would allow application logic to be run on server CPUs, on client CPUs, or offloaded to SmartNICs, with placement decisions automatically handled by the framework in response to changing workloads and resources.

\paragraph{Scalability in Multi-tenant Environments.}
Data centers host many applications; a robust framework must handle the diverse, concurrent demands of co-located tenants. State-of-the-art approaches to co-locating offloads rely on standard inter-process isolation between tenant code, which limits scaling as process context switch latency and overhead interfere.

Some existing on-path approaches sidestep these limitations by using specialized hardware and privileged software. For example, on an on-path SmartNIC iPipe~\cite{ipipe} avoids context switch overheads by controlling a software-managed TLB. These optimizations don't work on SmartNIC platforms without this specialized hardware and where the SmartNIC runs a commodity OS kernel that doesn't expose control of low-level address translation hardware to applications.

\paragraph{Other, Non-SmartNIC Frameworks.}
Nu is a framework for specifying functions that are co-located with in-memory state~\cite{nu}, which are dynamically migrated together to load balance, which is similar to NAAM's adaptive placement. Unlike Nu, NAAM doesn't migrate in-memory heap state (yet), instead, it ships functions to the data they operate on or uses remote access.

Splinter and systems that build on it execute arbitrary code extensions embedded in a kernel-bypass key-value store and dynamically shift code execution between server and client CPU cores based on CPU congestion~\cite{splinter,asfp,kayak}. In Splinter, a function cannot be migrated without restarting it, and it does not support execution across heterogeneous CPU architectures. These systems address questions about where to place execution, but, unlike NAAM, these placement decisions have to be made using software-based dispatchers rather than reconfiguring the hardware load-spreading features of NICs, and they don't use SmartNICs.  AIFM is a far-memory system that uses a limited set of pre-defined operations that are accelerated via running at remote memory nodes~\cite{aifm}.



\vspace{1eX}

Next, we describe NAAM's approach for developing efficient application-defined remote memory operations that scale to hundreds of tenants with dynamic load balancing across machines and commodity off-path SmartNICs.

%% file: tex/design.tex
\section{NAAM Design}
\label{sec:design}

\begin{figure}[t]
    \centering
    \includegraphics[width=\columnwidth]{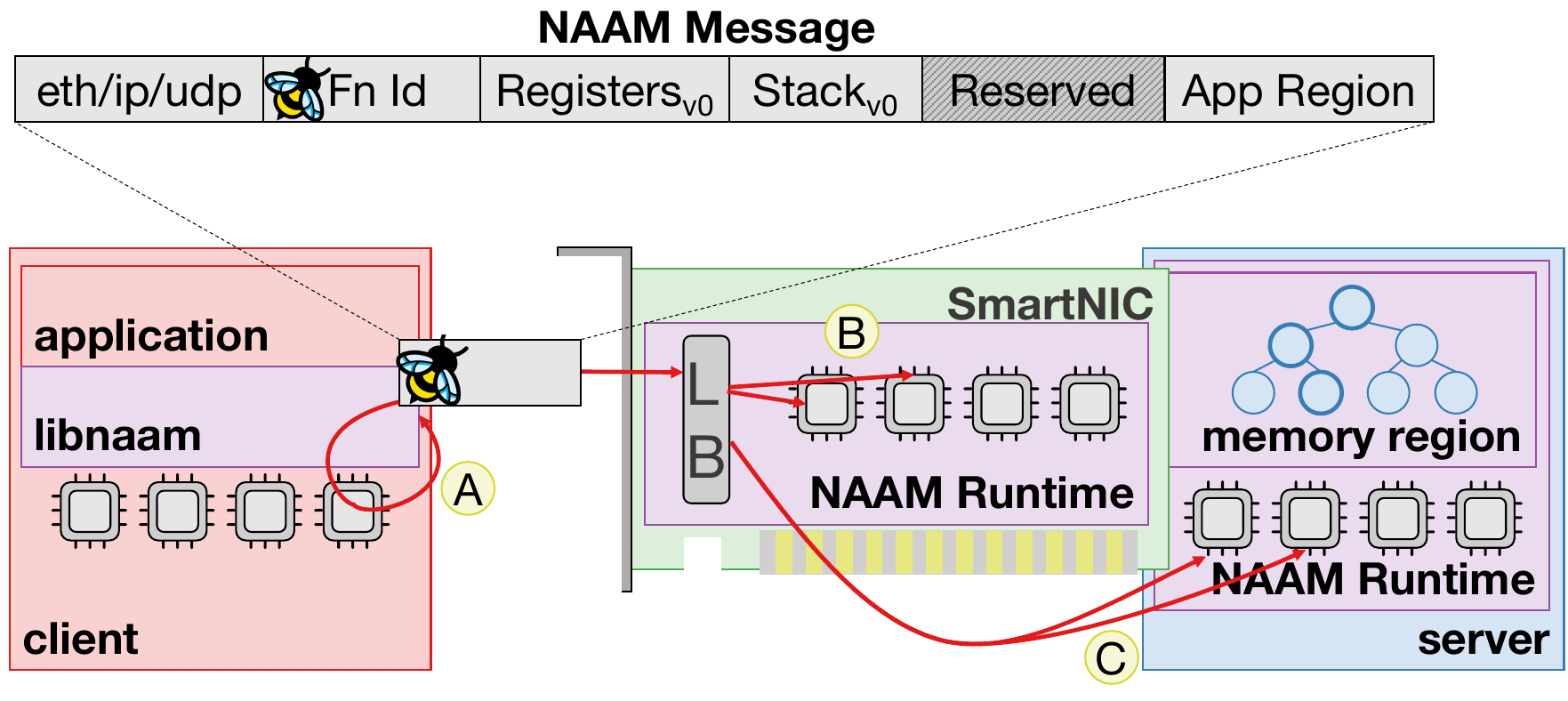}
    \caption{NAAM messages run application-defined functions on server-side state, and NAAM can adaptively run them on \tcirc{A} client CPUs, \tcirc{B} a server-side SmartNIC, or \tcirc{C} server-side CPUs.}
    \label{fig:concept}
\end{figure}



NAAM ({\sl{}network-accelerated active messages}) is a framework and runtime that allows applications to define custom remote memory operations whose processing is transparently offloaded between host and SmartNIC CPU cores. To overcome the limitations of existing frameworks that allow custom remote memory operations or that offload logic to SmartNICs, NAAM's design centers around five key goals\ga{I wonder if the goals shall be in sync with Table\ref{table:comparison} otherwise it might seem a bit disconnected?}:
\begin{description}[leftmargin=0.0em]
\item[Programmability.] NAAM remote memory operations are written in C and include arbitrary application code that can affect remote state. Messages can trigger simple RDMA-like operations, or they can perform complex operations like hash table or B-tree lookups. Applications use a standard RDMA-like interface in NAAM code to access state stored at servers.
\item[Efficiency.] NAAM message code
is compiled to eBPF bytecode and just-in-time compiled for each CPU architecture the message is processed on. This ensures that even application-defined logic in the messages runs at near-native speed whether running on a host x86 CPU or on a SmartNIC. We use eBPF due to its portability, security, isolation, and its proven support for a wide variety of network applications.
\item[Transparent Function Movement.] eBPF ensures that each NAAM message's associated code is portable; NAAM code is written once and can run anywhere. NAAM also ensures that each message captures the whole state of its running code. This allows load and processing to be redirected by simply redirecting packets. NAAM's novel eBPF extensions let it efficiently suspend and resume code (\S\ref{sec:udma-ops}).
\item[Non-blocking Memory Access.] Message code is inexpensive to suspend and messages are self-contained, so NAAM provides a novel RDMA-like interface to remote memory that avoids blocking, waiting, or synchronizing on data access. 
When remote data is needed, NAAM suspends execution of the message's code, forwards it to the location where the data resides, and resumes it there.
Messages run in a non-blocking fashion at each location until they are forwarded to fetch data, leading to a natural and efficient implementation of NAAM as a software switch. 
NAAM uses multicore and kernel bypass to efficiently fetch message batches, processes them in FIFO order, and forwards them to the data they wish to access.
\item[Multi-tenant Scaling.] Many applications may define custom operations. NAAM uses eBPF to isolate them from one another rather than hardware paging mechanisms, which are costly to reconfigure. This lets NAAM scale to orders of magnitude more custom operation types than existing approaches.
\end{description}

Figure~\ref{fig:concept} shows how NAAM messages are processed. Due to the portability of NAAM functions, each message's function can be executed anywhere the message encounters idle compute along the path despite the differences between the platforms. This includes \tcirc{A} client CPU cores, \tcirc{B} server CPU cores, or \tcirc{C} the SmartNICs that might sit in between them.

With the help of hardware flow steering, NAAM messages can be dynamically redirected to idle compute for the same cost as a simple network packet. This allows NAAM functions to adapt and provide high performance in the face of changing conditions in the data center environment, including host CPU, SmartNIC CPU, and network utilization even down to sub-second timescales (\S\ref{sec:monitoring}).


\subsection{NAAM Architecture}

\begin{figure}[t]
    \centering
    \includegraphics[width=\columnwidth]{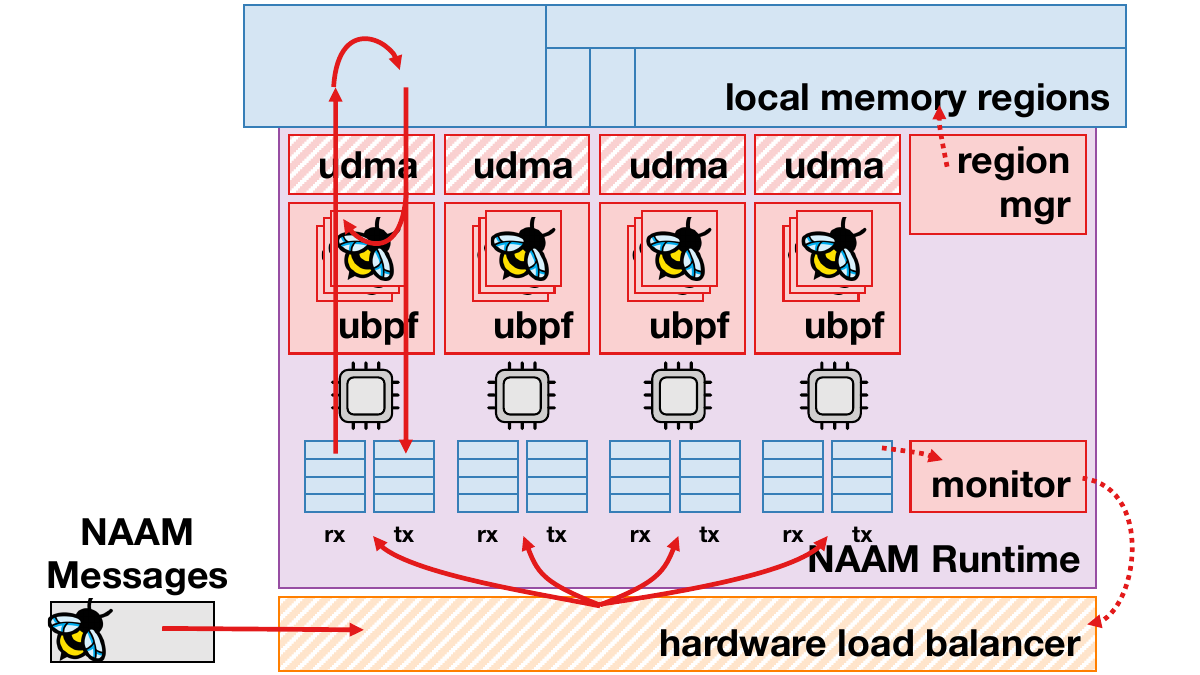}
    \caption{NAAM runs as a software switch on host or SmartNIC cores. NIC hardware balances NAAM messages across receive queues, one per device CPU core. Messages are processed by NAAM's uBPF module, which runs the JIT-compiled eBPF code for each message and then forwards the message to the UDMA module if data access is needed. UDMA reads or writes the memory region, and then forwards the message back to uBPF to finish processing until the message's code completes.}
    \label{fig:arch}
\end{figure}

To use NAAM, applications link to a \code{libnaam} library. Servers run one instance of the NAAM runtime on their host CPUs and another on their SmartNIC (Figure~\ref{fig:concept}); these instances coordinate to decide how to split work between them. Each instance runs as a software switch that processes NAAM messages using DPDK and an augmented verison of the uBPF userspace eBPF runtime~\cite{ubpf} (Figure~\ref{fig:arch}).

The client library lets applications create {\sl memory regions} that hold application data in NAAM. Each memory region resides in host memory or SmartNIC memory. Clients also use this library to register NAAM code for their application, which they provide as an ELF file that contains an eBPF function whose signature matches the signature of Linux's XDP in-kernel packet processing hook. These functions can be expressed in arbitrary C code (or in other languages that target eBPF), and they can access registered NAAM memory regions (via {\sl UDMA}s, \S\ref{sec:udma-ops}). Each registered function receives a unique function ID and destination UDP port number. Registration runs an eBPF verifier over the function before installing it and JIT-compiling in each instance's uBPF module.

After registration, the client can send NAAM messages that are tagged with the ID and port of that function. Instances of the NAAM runtime that receive this message run the function associated with the message using their uBPF module. These functions are passed an XDP context that refers to the NAAM message, so the code associated with a NAAM function operates on and transforms the NAAM message itself.

Generally, if a client sends a NAAM message to a server, the NAAM runtime at the server's SmartNIC receives the message first. NAAM configures the NIC hardware load balancer to divide work among available CPU resources; this includes CPU cores both on the SmartNIC and the host CPU. Each CPU has a dedicated receive queue and processes NAAM messages independently of the other CPU cores.

Whenever a function running in a uBPF module requests data from a memory region, the function is suspended and its state is captured into the message. Then, the message is forwarded to a UDMA module on the server where the requested data resides, which reads or writes the requested data from the memory region. Afterward, the message is recirculated to NAAM's uBPF module to resume executing the function. Once the function completes, the resulting (transformed) message is transmitted back to the client via the core's transmit queue.

This design meets NAAM's key goals. It is programmable and efficient: NAAM messages contain arbitrary code JIT-compiled to the underlying architecture, eBPF's portability ensures NAAM can steer messages anywhere for execution, and its scheme for function migration allows NAAM cores to forward the message to the data it wants to access rather than holding the message while awaiting remote data. Next, we describe NAAM's memory regions and its UDMA operations in detail, then we describe its load balancing; finally, we describe NAAM's security model.


%

\subsection{Memory Regions}
\label{sec:mr}

NAAM memory regions are allocations of memory that can be accessed by NAAM messages to store application state. Each region has a unique ID and a fixed size, and it can reside in the host's local memory or on-NIC memory. Applications use a control-plane RPC to a NAAM service to create memory regions. Afterward, these regions are globally addressable by NAAM functions via their memory region ID and offset. Memory regions are implemented using shared memory so NAAM runtime instances can access them from any core. If a NAAM message is running on SmartNIC cores and it requests access to a memory region in the host CPU's memory, then the access is performed using DMA from the NIC.

\subsection{UDMA}
\label{sec:udma-ops}

NAAM messages may need to access server-side state wherever they run, which NAAM's {\sl universal direct memory access} (UDMA) interface provides.
All external state accesses by NAAM functions happen through eBPF helper functions that trigger UDMA operations, which have similar semantics to RDMA operations. However, unlike RDMA operations, UDMA operations are executed differently depending on where they are triggered. When running on the client, UDMA operations trigger remote DMA operations to the server; when running at the server's SmartNIC they trigger local DMA operations over PCIExpress; and when running at the server CPUs they manifest as local \texttt{memcpy}s. This ensures that all state accesses use one interface regardless of where the message's code executes while eliminating the overhead of full RDMA operations when possible. Hence, the same eBPF function can be used to access state regardless of where the function is executed, providing a unified programming model for writing remote memory applications. Stateful applications can be written once, compiled to eBPF, and run anywhere.

\subsubsection{UDMA Interface}



\begin{table}[t]
\centering
\footnotesize
\begin{tabular}{p{0.95\columnwidth}}
\toprule
\textbf{struct addr\_t} \{u8\, region\_id;\, u64\, offset\}; \apigap{}

\textbf{APP\_REGION}(void *msg) $\rightarrow$ (size\_t, void*) \\
\qquad Get offset and pointer to usable packet buffer region. \apigap{}



\textbf{UDMA}(xdp\_md *ctx,\, addr\_t\, dst,\, addr\_t\, src,\, u64\, len) $\rightarrow$ u8 \\
\qquad Copy len bytes from src to dst. Returns 0/1 on success/failure. \apigap{}


\textbf{UCAS}(xdp\_md *ctx,\, addr\_t\, dst, u32\, old,\, u32\, new) $\rightarrow$ u32 \\
\qquad Atomic compare-and-swap; *dst = new if *dst == old.\\
\qquad Returns *dst from before swap. \apigap{}

\textbf{UFAA}(xdp\_md *ctx,\, addr\_t\, dst,\, u32\, val) $\rightarrow$ u32 \\
\qquad Atomic add; *dst += val. Returns *dst from before add. \\ \bottomrule
\end{tabular}
\caption{API for Network-accelerated Active Messages}
\label{table:api}
\end{table}

Table~\ref{table:api} lists the interface that NAAM exposes to NAAM functions to perform UDMA operations.

\paragraph{Read/write.} A NAAM message uses \code{UDMA(ctx, src, dst, len)} to move data to/from a memory region and its message buffer (Table~\ref{table:api}). \code{src} and \code{dst} each specify a memory region id and an offset into it. Memory region 0 is special; it always refers to the message's own buffer. Commonly, UDMA operations specify a destination with memory region 0 with offset \code{APP_REGION(msg).offset}; this copies bytes out from the source memory region stored at the host, letting the message code access it via its buffer, similar to an RDMA READ. Flipping the source and destination moves data from the message to the host memory region, similar to an RDMA WRITE.

\paragraph{Atomic Operations.}
\code{UCAS} and \code{UFAA} perform atomic compare-and-swap and atomic fetch-and-add operations on a 32-bit value in a memory region. These operations execute native atomic operations on the host memory if the function runs on the host. Otherwise, they perform atomic operations over DMA/RDMA if the function runs on the NIC. This only works if the NIC supports either atomic DMA operations or RDMA atomics. If one of these isn't supported, atomic operations are forwarded to the UDMA module at the correct host CPU, leveraging NAAM's location independence.


\begin{lstlisting}[float=t, language=C, caption={Linked list traversal as a NAAM function.}, label={lst:naam-linked-list}]
typedef struct __attribute__((__packed__)) {
	u32 val;			// Value in the linked list.
	u32 next_off;	// Next node offset in memory region.
} llnode_t;

// Returns last node in linked list stored in a memory region.
SECTION("xdp")
u64 prog(struct xdp_md *ctx) {
  // Prove to the verifier no out-of-bounds accesses.
  if (ctx->data + offsetof(req_pkt_t, ctx->data) + sizeof(llnode_t) > ctx->end) return 1;

  // Refers to the head node at the start of memory region 1.
	addr_t src{1, 0};
	// Refers to the app-usable region of NAAM message buffer.
	addr_t dst{0, APP_REGION(msg).offset};

	for (u32 i = 0; i < 16; i++) { // 16 max len bounds loop.
		// Read node from the memory region to message buffer.
		u8 ret = UDMA(ctx, dst, src, sizeof(llnode_t));
		if (ret == 1) return 1; // Return failure.
		
		// Fetched node is now in the message buffer.
		llnode_t *node = (llnode_t *)APP_REGION(msg).addr;
		if (node->next_off == -1) break; // Found end.
		
		// Next node in the list.
		src.offset = node->next_off;
	}
	return 0;
}
\end{lstlisting}

\subsubsection{UDMA Example} Listing~\ref{lst:naam-linked-list} shows example NAAM message code that implements traversal of a linked list stored in memory region 1 on a host. On line~13, offset 0 in the memory region 1 is set as the source address, which is the address of the head node in the linked list. On line~15, the message-usable region of the buffer is set as the destination. Line~19 sets up a UDMA descriptor in the message buffer with \code{src} and \code{dst} for length 8, which is the size of each list node. Once \code{UDMA()} creates the descriptor, it forces the code to yield back to the NAAM runtime, which forwards the message to the UDMA module. The UDMA module copies the requested data into the message, and it forwards the message to the uBPF module, which resumes execution of the message code where it left off. The code extracts the next node's offset (Line~27), and this repeats until the end of the list has been reached (Line~24), or \code{MAX_LEN} iterations have occurred.

This code is compiled to eBPF bytecode and sent to all the locations where the function will be executed. \code{UDMA()} will perform local memory accesses if the function runs in the host or DMA if the function runs on the NIC. If the code runs at the client \code{UDMA()} will fetch nodes from the host memory.

\begin{figure}[t]
    \centering
    \includegraphics[width=\columnwidth]{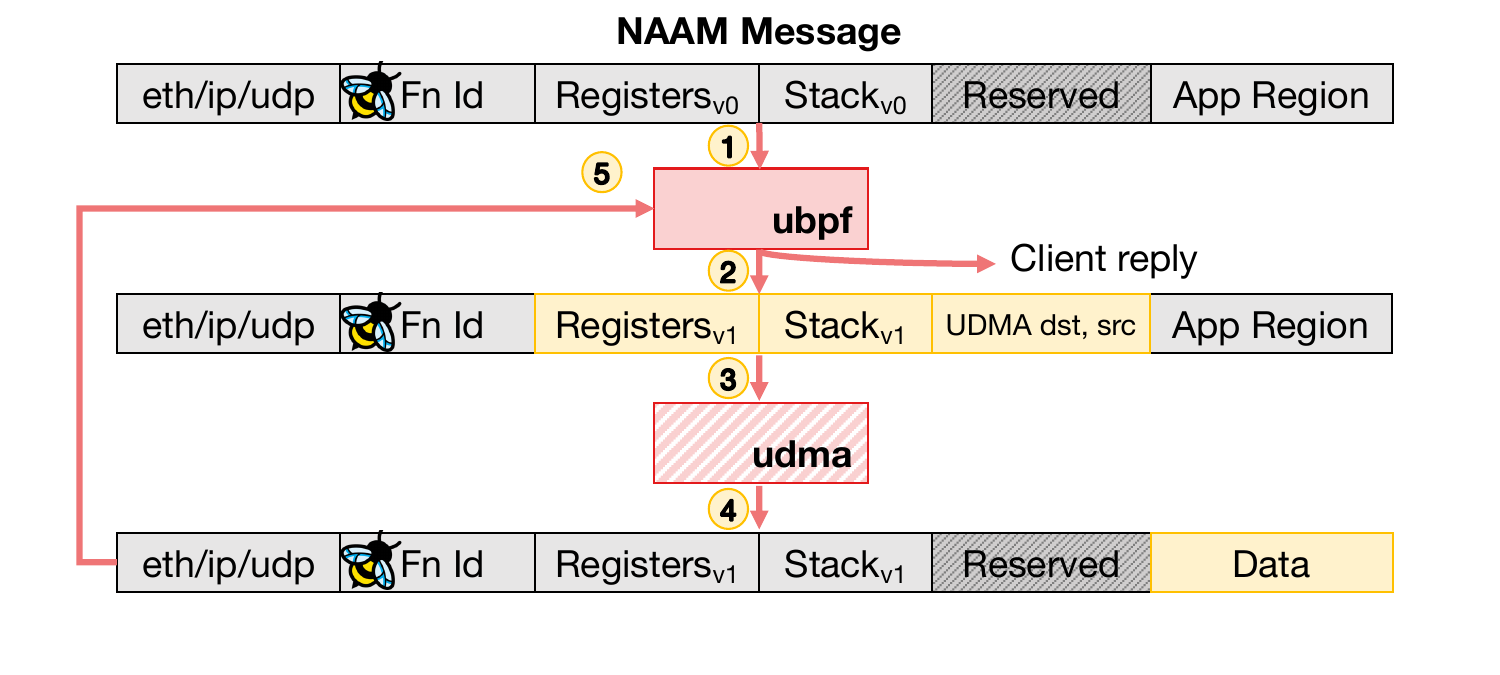}
		\caption{\tcirc{1} Clients send NAAM messages. The NIC forwards the message to where it will be executed. \tcirc{2} The uBPF module executes it; it forwards the message as a reply to the client if the function exits, \tcirc{3} otherwise, on UDMA operations the uBPF module stores the function state into the message, sets up a UDMA descriptor and forwards the message to a UDMA module. \tcirc{4} After the UDMA module fills the message with requested data, it forwards the message back to the uBPF module. \tcirc{5} The uBPF module uses the function state in the message to resume at the next instruction.}
    \label{fig:naam-msg-flow}
\end{figure}

\subsubsection{Non-Blocking UDMA Execution}

A key challenge with UDMA operations is that they can vary in response time depending on where NAAM functions are run. Stalling the processing of a NAAM message to await remote operations could result in head-of-line blocking or high synchronization costs to suspend operation on one NAAM message to start processing another. Having deep queues of packets awaiting remote operations is costly.

To avoid this, NAAM adds new eBPF functionality to support functions that cooperatively yield while saving register and stack state. NAAM messages store this register and stack state directly in the message's packet buffer~(Figure~\ref{fig:naam-msg-flow}, \tcirc{1}).
Whenever a NAAM message calls \code{UDMA()} to access a memory region, a descriptor is added to the message buffer indicating what data should be read or written by the operation (\tcirc{2}). Then, the runtime saves the running state of the function into the message, and the whole message is sent to a UDMA module where the region resides (\tcirc{3}). The UDMA module copies the requested data based on the descriptor in the message (\tcirc{4}). Finally,  the UDMA module recirculates the message to a uBPF module (generally on the node where the UDMA completed) (\tcirc{5}), and the function can be seamlessly resumed after the requested data has been copied into the NAAM message.

In some circumstances, NAAM's uBPF and UDMA module can be separated by the network.
The function can be resumed on a different machine or different CPU architecture. Hence, unlike most approaches where suspended code must await and synchronize with remote responses for data, NAAM messages are self-contained. This allows them to be processed from FIFO queues without introducing stalls or extra overhead. On remote access, execution is suspended into the NAAM message, and the entire message can be forwarded to the data and resumed there (or elsewhere).

For example, NAAM also supports running functions at the client's CPUs. In this case, the uBPF module runs as part of the client's \code{libnaam} code, which then forwards messages to a remote server's UDMA module in order to perform remote state accesses. It is also possible that a function running on one device wishes to access a memory region stored on another device, in which case the NAAM message must be forwarded between the uBPF module on one machine and the UDMA module on another.

When a UDMA module on a SmartNIC must access state stored in a memory region stored in host memory,
it uses local DMA over PCIExpress instead of \texttt{memcpy}. On the BlueField-2, DMA operations have roughly 3.5~\textmu{}s latency, so NAAM takes advantage of this context switch mechanism to pipeline them. While one batch of NAAM messages is awaiting the completion of DMA operations in the UDMA module, the uBPF module executes the functions for another batch of NAAM messages. This helps hide the cost of DMAs that functions trigger from a SmartNIC to access host memory.

\paragraph{Cooperative Yields.}
Each NAAM message buffer contains network headers followed by program state, consisting of a stack, register values, and a program counter (Figure~\ref{fig:naam-msg-flow}). The rest of the buffer is accessible to the program to use as inputs or scratch space. One register always holds a pointer to an XDP context, which contains a pointer to the message buffer; another register holds the uBPF VM stack pointer.

When the VM starts a function, it loads the VM registers from the buffer, loads the buffer's stack area into the stack pointer register, and executes the function from the first instruction. Directly using the stack from the message buffer eliminates the need to copy the stack when suspending and resuming the program. This reduces state copy overhead in yield points when the function is suspended and resumed.

When \code{UDMA()} is called, the active program counter and register values are copied into the message buffer, including the stack pointer, so nothing more has to be done for the stack. The parameters to \code{UDMA()} are used to form a UDMA descriptor in the buffer. Then, the message buffer is forwarded to the UDMA module. After the UDMA completes, this or another uBPF module can resume the function by restoring the registers and stack pointer. For JIT-compiled eBPF, we append context-restoring code to the start of the program, and we inject context-saving code at every yield point.


On resuming from these context switches, NAAM can run the program on any uBPF module. For example, part of a NAAM program can run on the SmartNIC, and then it can be resumed at the host, even if their CPU architectures differ. Furthermore, since UDMA parameters are stored inside the packet buffer, uBPF modules and UDMA modules can live across machine boundaries from each other. For example, the uBPF module can run eBPF code in the client, and the DMA module can run in the server SmartNIC, which would give RDMA-like execution to the function. Alternatively, a uBPF module may run together with a DMA module, which would give the program RPC-like execution, where the program and the data it accesses are local and co-located.
\rs{Kill this now that UDMA Module sec above says something similar?}


\subsection{Hardware-Assisted Load Balancing}

To enable dynamic placement of NAAM functions, NAAM messages must be quickly and efficiently steered to the locations where they will be executed. Commodity off-path SmartNICs support hardware flow steering to redirect traffic from the NIC physical port to host CPU cores, to SmartNIC cores, or both. Openflow~\cite{openflow} rules can be installed to program the flow steering hardware. Initially, low-priority rules are installed on the hardware to redirect all the traffic to the SmartNIC cores. Then, if the SmartNIC cores are overloaded, new rules are installed with a higher priority to redirect a fraction of the traffic to the host. Redirecting traffic shifts load from one place to another as it moves function execution from one place to another. Traffic is matched based on the UDP source port numbers to divide it up. For example, consider ten flows with ten different source port numbers coming from the clients and going to the SmartNIC. If the SmartNIC is overloaded, an OpenFlow rule will be installed in the flow steering hardware to send one of these flows to the host. This moves 10\% of the traffic to the host and removes 10\% load from the SmartNIC cores. On the other hand, if the SmartNIC cores are underloaded, this rule can be deleted to move 10\% of the traffic from host CPU cores to SmartNIC CPU cores. \code{libnaam} at each client chooses UDP source port numbers in a fashion that helps the matching for hardware flow steering.

\subsection{Resource Monitoring and Load Shifting}
\label{sec:monitoring}

NAAM provides system designers with a framework for implementing policies to automatically move functions based on workload requirements and system resource usage.

To help with this, NAAM tracks the time messages spend in NIC receive queues using
NIC hardware timestamping.
Each time a NAAM message is received, it stores the NIC's clock cycle count in a metadata field in the message.
The NIC and the CPU clock have different frequencies. On NAAM initialization, the difference between the clocks is determined. Each time the clocks are used, NAAM compensates for the frequency difference in order to determine message queuing delay.
Reading the timestamp for each message is costly, so NAAM reads the timestamp of one message from each batch.

NAAM calculates the average queue time in 10~ms windows. If three consecutive windows out of five are above a threshold, then NAAM decides that the host is overloaded, and it installs rules in the NIC flow steering hardware to move a fraction of the load to the NIC. We have used this mechanism to detect whether there is interference from another application in a latency-critical NAAM application. When there is interference, NAAM function execution is moved to SmartNIC cores.

NAAM also monitors packet loss which can be used as a signal for reconfiguring how messages are steered between host and SmartNIC cores. To give NAAM's monitors a global view, a monitoring daemon on the host CPUs pushes statistics to the NAAM monitoring daemon on the SmartNIC which makes reconfiguration decisions.


\subsection{Security}
\label{sec:security}

NAAM functions are associated with specific memory regions, and all accesses that an invocation of a function makes are restricted to target (a) the invocation's private register file or stack (stored in its message buffer) or (b) an authorized memory region for that function via UDMA operations. To achieve this, NAAM relies on five key properties.

\paragraph{Memory Safety.} First, NAAM relies on eBPF to ensure running NAAM function code can only access the function's VM state or call UDMA helper functions.
The Linux eBPF runtime~\cite{kernel-ebpf} has a verifier that statically checks the memory and type safety of a program to ensure it can only access its own stack and register state and that it respects the expected calling conventions for trusted kernel helper functions. Since the Linux eBPF verifier isn't easy to extract from the Linux kernel, NAAM provides these guarantees using the PREVAIL eBPF verifier~\cite{prevail-verifier}, which it runs in userspace at function registration time. PREVAIL works unchanged on NAAM messages that contain no yield points; however, yield points create a special challenge. Specifically, a function invocation's stack and register file state can move (in memory or even between nodes) in between the yield and resume, creating challenges for pointers that are embedded in the saved stack and registers. We discuss how our extensions to PREVAIL's verifier solve this issue safely and efficiently in Section~\ref{sec:impl}.

\paragraph{Memory Region Allow-Lists.} Second, each UDMA engine keeps an allow-list of functions that can access each memory region. Since functions cannot tamper with state outside of their VM and memory regions, they cannot tamper with the allow-list or the function ID associated with their execution.

\paragraph{VM State Initialization.} Third, NAAM uses a trusted module that maps UDP destination ports in client messages to function IDs and ensures that the message buffer contains an initialized and ``zeroed'' state for the message's function execution including its registers and stack. This ensures each client request message first enters its NAAM function in the same state that the verifier assumed when statically checking the safety of the NAAM code. NAAM's eBPF runtime ensures that functions cannot directly tamper with the regions of their message buffer that contain VM state (e.g.\ via pointers or UDMAs), which ensures the verifier's assumptions are upheld. NAAM performs no special authentication of incoming messages targeting NAAM functions' associated UDP port numbers; any message targeting a function's associated UDP port number is allowed to execute after safe VM state initialization. Hence, if application-level access control policies are needed, NAAM function code is expected to enforce them.

\paragraph{Migration Integrity.} Fourth, whenever a NAAM function's execution is suspended and forwarded to another node to continue execution, the NAAM runtime sends the message to a special destination UDP port number on the target node. This acts as a flag indicating that the VM state was generated by another NAAM node and that it represents a valid suspended state for the function associated with the message. Clients should not generate messages with this destination UDP port number, since they might craft arbitrary saved register and stack state that would violate the verifier's assumptions and could lead to unsafe execution. One way to enforce this is to trust network source and destination addresses and to have NAAM nodes reject messages targeted to this specific UDP port number unless they are generated by another known NAAM node. Our current implementation does not make this check; instead, we trust clients not to send messages directly to this UDP port number. Since NAAM functions can modify message headers, they could craft packets that target this UDP port number. To prevent this, NAAM checks the UDP destination port after each function yield and it discards any message that has this special UDP port number as its destination.

\paragraph{Network Trust Model.} Finally, NAAM trusts the network for any secrecy and integrity properties that the operator desires. For example, if the network or a node attached to the network modifies the function ID, destination port number, or VM state of any NAAM message, NAAM's safety guarantees breakdown. In networks without this trust, connections between the NAAM nodes and between their clients and the nodes would need to be protected e.g.\ via TLS.

\section{Implementation}
\label{sec:impl}

NAAM's self-contained and stateless messages leads to a natural implementation of the NAAM runtime as a platform-agnostic virtual switch where these stateless messages can flow between different operational components of the switch. NAAM is implemented on top of the BESS virtual switch~\cite{bess}. BESS is kernel bypassed, high-performance, easily extensible, and modular. Each BESS module implements a specific NIC feature and can send/receive packets from other modules. Pipelines are created by composing these modules. We have added the uBPF module to execute eBPF bytecode and UDMA modules for performing location-independent memory accesses. Additionally, we have ported BESS to the ARM architecture so that it can be run on both x86 and ARM platforms to make it a platform-agnostic virtual switch. The uBPF module uses uBPF~\cite{ubpf}, a userspace implementation of an eBPF virtual machine for executing eBPF bytecode. We have modified the interpreter and the x86 and ARM JIT compiler to support cooperative yield via helper function calls that save/restore contexts to/from NAAM messages.

\paragraph{Cooperative yield.}
To implement cooperative yield, we augmented uBPF's interpreter and JIT-compiler. Each yield returns twice. The first is when \code{UDMA()} is called. An eBPF \code{ret} instruction is executed that passes a return code (via \code{r0}) to the uBPF module to signal that the function has yielded and that the message should be forwarded to the UDMA module. Then, when the message comes back from the UDMA module, after restoring the state from the message, the status of the UDMA operation is assigned to \code{r0}, and the function resumes after the previous \code{ret}. From the running function's perspective, this final value of \code{r0} is the return value of the \code{UDMA()} call.

Implementing this on the eBPF interpreter is easy as the full state of the running function is available; however, implementing this in the JIT compiler is more challenging as access to the program state is limited. For example, to save or restore program state, the JIT-compiled code needs access to the message. By convention, \code{r1} contains a pointer to the context, which contains a pointer to the message, but this register is also one of the five registers used to pass parameters to eBPF helper functions. So, before calling a helper function, the eBPF compiler always moves context pointer to another register. To solve this problem, the first argument to the yielding call is always a pointer to the context; this ensures that \code{r1} will hold a pointer to the context upon yield.

\paragraph{Message Buffer Relocation.}
One challenge with resuming yielded functions is that the location of the message buffer might change before and after a yield point. For example, a function could begin executing on one machine; after a yield, the NAAM message could resume on another machine where the location of the message buffer has changed in the address space of the NAAM runtime. Without care, this could lead to incorrect function behavior, and it could allow functions to access and corrupt arbitrary state in the NAAM runtime.

To solve this issue, NAAM uses a combination of extensions to its eBPF verification and to its runtime. At a high level, it uses the verifier to assess which registers and stack locations contain pointers to locations in the message buffer. NAAM function compilation takes this information as input to generate correct relocation code; on yield and resume, the code injected during compilation informs the runtime of which addresses need relocation to ensure they refer to the correct, inbound offsets in the new message buffer.

In more detail, the yielding UDMA calls in Table~\ref{table:api} have a corresponding internal implementation helper function that takes an extra 64-bit vector as an argument. This vector indicates which saved registers and stack locations contain pointers that need relocation on resume. Only four registers from \code{r6}-\code{r9} need to be checked, since they are the callee save registers; the other registers can be zeroed out when resuming at a yield point. The four least significant bits of the vector correspond to these registers; the remaining 60 bits correspond to the first 60 naturally-aligned eBPF stack locations, which are the only stack locations that can hold pointers. However, in practice there are 64 stack locations (in eBPF's standard 512~B stack); if our verifier encounters a message buffer pointer in the last 4 potential stack locations it rejects the function. This is safe, but it disallows some otherwise safe functions that yield while holding pointers on their nearly full stack. This could be solved by passing a longer vector to each yielding call, but no function that we wrote in the course of this work spilled a message buffer pointer to \emph{any} location on the stack, so this hasn't been a concern thus far.

If a bit is set in the bit vector that means the corresponding register or stack location contains a pointer to the message buffer or stack and it needs relocation. Running NAAM's PREVAIL-based verifier outputs information indicating the needed register and stack information for each yield. A script parses this information and constructs the vector for each UDMA call. Then the NAAM function is recompiled with the correct bit vector for each UDMA call. At runtime, NAAM uses this vector together with the saved \code{r1} which contains the old message buffer location to do relocation as follows: \code{state[i] = new_r1 + (state[i] - r1)}
 where \code{state[i]} contains the corresponding register or stack location. Then, the new message buffer address is assigned to \code{r1}, and the function is resumed from the next instruction.


\paragraph{Atomic UDMA operations.}
To perform atomic operations from the NIC on host memory, the NIC must support atomic DMA operations or RDMA atomics. The BlueField-2 DOCA library does not expose atomic DMA operations~\cite{doca-dma}. RDMA atomics support two modes~\cite{rdma-manual}; \code{IBV_ATOMIC_HCA} guarantees atomicity on-device, which is insufficient to synchronize with host CPU operations, and
\code{IBV_ATOMIC_GLOB}, which synchronizes across host CPU and devices. The BlueField-2 supports \code{IBV_ATOMIC_GLOB}, but we were unable to get it to work despite using it on a supported PCIe configuration.
To solve this, \code{UCAS} and \code{UFAA} operations are forwarded to the UDMA module on the host with the corresponding memory region.

\paragraph{NIC flow steering rules.}

NAAM uses NIC hardware flow steering to shift computation between host and SmartNIC resources.
To support fine-grained compute movement, we considered three approaches to configure the flow steering rules. The first uses OpenFlow weighted group bucket rules~\cite{openflow-spec} which allow NAAM to directly control the fraction of traffic sent to the host and NIC.
The rule can be updated to adjust the fraction. The second uses OpenFlow range-based rules~\cite{openflow-spec}, where a range of flows can be specified to split traffic. By controlling the range, traffic can be split between the host and NIC. This requires updating two rules to update the split. Our experiments showed significant packet loss with both approaches since these rules trigger software paths on the SmartNIC. Hence, NAAM uses a one-rule-per-flow approach; this approach moves traffic in sufficient enough granularity with a modest number of flows. With ten rules and ten flows, compute can be moved at 10\% granularity.

%% file: tex/evaluation.tex
\section{Evaluation}
\label{sec:eval}

To evaluate NAAM, we use both synthetic applications and real-world workloads -- MICA, an efficient hash table~\cite{lim2014mica} and a B-Tree similar to Cell~\cite{cell}. We seek to answer these questions in evaluating NAAM:

\begin{enumerate}[noitemsep,leftmargin=1.5em]
	\item How does NAAM compare with the state-of-the-art SmartNIC offloading frameworks (iPipe)~\cite{ipipe}?
    \item Does dynamic offloading allow a server to scale beyond the throughput it can achieve with CPU cores alone?
	\item Can dynamic offloading react quickly enough in response to load changes to avoid bottlenecks?
	\item Can NAAM mitigate CPU interference on host CPU cores by quickly detecting and reacting to interference?
	\item What is the trade-off for running NAAM functions in different locations (client side, server side, and server side with automatic SmartNIC offload enabled)?
    \item Can NAAM prevent crashes due to functions with bugs?
	\item What is the cost that eBPF adds to NAAM functions compared to using natively compiled code?
    \item How does NAAM compare with RDMA on applications?
\end{enumerate}

\paragraph{Experimental Setup.}
All experiments use two machines, one acting as a client and another acting as a server. The client machine has two 12-core Intel Xeon Silver 4310 CPUs, 128 GB RAM, and an NVIDIA ConnectX-6 100~Gbps NIC. The server has the same CPUs, 256 GB RAM, and an NVIDIA BlueField-2 200~Gbps SmartNIC. Both NICs use a PCIe $\times$16 link. The SmartNIC runs firmware v24.37.1014, DPU BSP v3.9.3 as the OS, and NVIDIA DOCA library v1.5.0. It is set to run in SmartNIC mode with hardware offloading enabled. This reduces the latency since, initially, all flows go to its ARM cores, but subsequent packets are directly passed from the hardware switch to the destination cores without routing through the ARM cores first.  One port on the SmartNIC is connected directly via Ethernet to a 100~Gbps port on the client machine's NIC without an intervening switch.


The client uses DPDK~v20.11~LTS to generate load in an open loop.
All SmartNIC experiments process messages on six of the eight SmartNIC cores; the other two cores are used for NAAM control plane services.

\subsection{Scaling to Many Functions}

\begin{figure}[t]
\centering
\includegraphics[width=0.8\columnwidth]{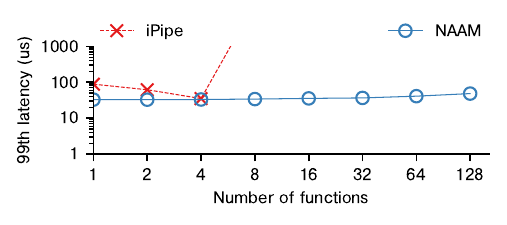}
\caption{Multi-tenancy scaling for iPipe and NAAM.}
\label{fig:ipipe}
\end{figure}

Multi-tenancy is important in data centers since many co-located applications run together on shared hardware. Since NAAM's eBPF-based approach to function isolation doesn't require special hardware support, it scales independently of CPU cores, address-space identifiers, or other hardware-limited resources. This distinguishes NAAM's approach from other state-of-the-art SmartNIC offloading approaches.

For example, iPipe is a state-of-the-art actor-based approach to offloading functions to SmartNICs. Its actor-based model is attractive for NAAM-like applications since actors can be small, and the state they act on can be explicitly contained in each actor. However, iPipe's inter-actor isolation model relies on specialized hardware features specific to Marvell's Liquid~IO~II on-path SmartNIC. Specifically, iPipe establishes separate TLB entries for each actor in a software-managed TLB on the SmartNIC's MIPS cores. This allows each SmartNIC core to receive a packet and then run code on behalf of any actor without the high cost of address space switching required in Linux on conventional CPU architectures like Intel and ARM. NAAM's approach avoids the cost of address space switching by using eBPF for isolation rather than unconventional address translation hardware.

The BlueField-2's ARM CPUs only support conventional page table switching for address space isolation, so iPipe's actors must be run as separate processes as the authors suggest~\cite{ipipe}. Processes' conflation of threads and address spaces requires context switching between actors that share a core; this requires invoking the Linux kernel scheduler on the BlueField-2, which has a high cost. The result is that iPipe's performance breaks down if there are more actors than cores on the NIC since frequent context switching is needed.

Figure~\ref{fig:ipipe} shows this effect. In this experiment, the client sends a low rate of MICA hash table lookup requests to the server; each request is processed entirely at the SmartNIC. The code needed to handle these requests is run by NAAM and our iPipe setup with an increasing number of functions and actors, respectively. The separate buffer pool needed for each iPipe process makes it difficult to run with more than four processes, so these results limit both frameworks to four cores; without this restriction, iPipe could scale to eight actors but its performance would still collapse at 16 actors when there are more actors than SmartNIC CPU cores.

Because NAAM relies on eBPF, many different functions can run on the same thread at low cost. Hence, NAAM's hardware load balancing at the SmartNIC routes NAAM messages to cores randomly; the core that receives a request runs the associated JIT-ed eBPF function and sends back a response. The results show that NAAM can run a mix of 128 different functions at the same request rate with only a 47.3\% increase in 99\textsuperscript{th}-percentile response latency compared to when all requests target the same function.

The ``iPipe'' line shows an approach similar to what iPipe suggests for using off-path SmartNICs like the BlueField-2; the original iPipe code only supports on-path SmartNICs. In this set up each actor is run as a separate process. We use the SmartNIC's hardware load balancer to route incoming packets to the correct actor; each process receives packets from a separate queue, which avoids synchronization costs between the actors to dispatch requests among actors. As the results show, iPipe's approach works well when only one actor is scheduled per available SmartNIC core, but when there are more inter-isolated actors, the performance breaks down as the kernel scheduler must begin timeslicing the actor processes on cores. This slowdown results in tens of milliseconds of response latency even at low load. Results under higher load would be worse since this context switching also leads to a large throughput loss at higher request rates.

In sum, the state-of-the-art approaches to SmartNIC offloading provide good performance when used to shift remote-memory-style operations to SmartNICs, but CPU paging hardware and context switch overheads prevent them from scaling to more than a handful of inter-isolated functions.

\subsection{Dynamic Load Shifting}

\begin{figure}[t]
\centering
\includegraphics[width=0.8\columnwidth]{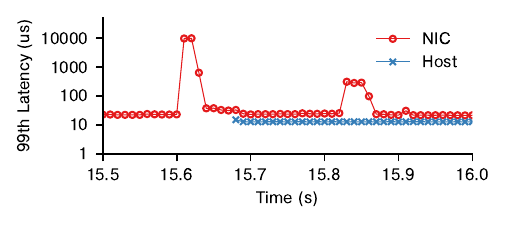}
\caption{Impact of hardware flow steering reconfiguration.}
\label{fig:flow-steering}
\end{figure}

To efficiently shift load, NAAM must reconfigure the NIC's hardware flow steering rules based on CPU load and queuing; implementing this in software would be too costly, and it would use nearly all of the available SmartNIC CPUs.
To redirect traffic in real time in response to the changing CPU load, NAAM uses a monitoring script on the SmartNIC that detects packet loss at the host and the NIC. The script reconfigures hardware flow steering rules to shift load as needed.

Reconfiguring hardware rules could delay or disrupt ongoing traffic.
To measure this disruption, we run an experiment where new load balancing rules are installed while under steady load. Figure~\ref{fig:flow-steering} shows the response time of messages that perform a 16~B UDMA writes to server memory over time. Initially, all messages are processed at the SmartNIC; at 15~s a rule is installed to shift 10\% of the traffic to the host. The rule takes effect, and queues build over about a 50~ms window (at 15.6~s); after less than 100~ms messages begin processing at host CPU cores, restoring low response times. The shift is fast enough that it induces no loss.

\begin{figure}[t]
\centering
\includegraphics[width=0.8\columnwidth]{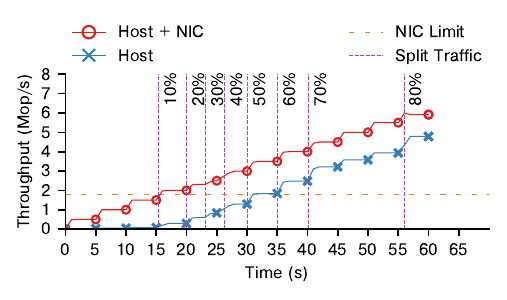}
\caption{Dynamic load shifts prevent SmartNIC overload.}
\label{fig:load-shift}
\end{figure}

To show that NAAM can continuously shift load between SmartNIC and host CPU cores online as needed, we run an experiment that slowly increases client-offered load from 500,000 MICA ops/s every 5~seconds. The results are shown in Figure~\ref{fig:load-shift}.
The circle data points show the total client-observed throughput; the cross data points show the throughput that is due to the ops processed by the server host CPU rather than its SmartNIC. Initially, all the responses are from the SmartNIC.
At 15~s the offered 2M~ops/s exceeds the throughput that the SmartNIC can sustain; 
NAAM detects the resulting packet loss and shifts 10\% of the traffic to the host.
As the offered load increases, more traffic is shifted to host CPU cores. The graph shows that the system scales beyond the 1.8~messages/s that the SmartNIC can process by dynamically shifting load to the host when it is overloaded.

\subsection{Mitigating Host CPU Interference}

\begin{figure}[t]
\centering
\subfloat[Host CPU; no interference.]{%
    \centering\includegraphics[width=0.8\columnwidth]{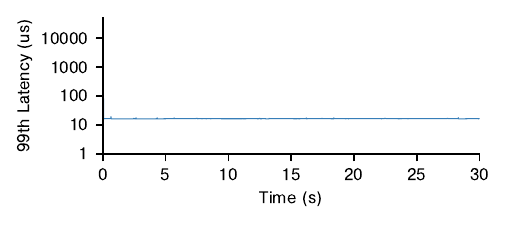}%
    \label{fig:no-interf}%
}
\vspace{-2eX}
\subfloat[Host CPU; interference from 10~to~20~s; monitoring off.]{%
    \centering\includegraphics[width=0.8\columnwidth]{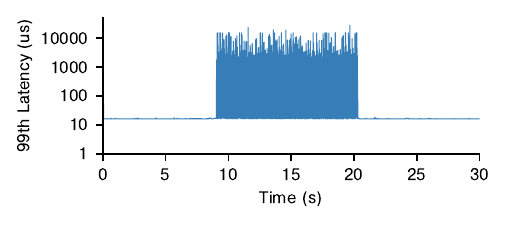}%
    \label{fig:interf}%
}
\vspace{-2eX}
\subfloat[Adaptive; interference from 10~to~20~s; monitoring on]{%
    \centering\includegraphics[width=0.8\columnwidth]{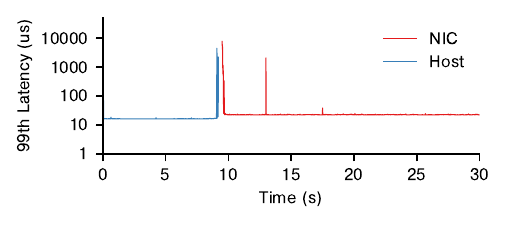}%
    \label{fig:interf-shift}%
}

\caption{NAAM shifts load to avoid host CPU contention.}
\label{fig:interf-full}
\end{figure}
Latency-critical networked applications often avoid disruptions due to kernel thread scheduling decisions by dedicating CPU cores to applications. Even so, these disruptions can occur, especially when a server's CPU becomes oversubscribed. For example, a Spark job might be scheduled on a CPU core ordinarily used for processing requests to a networked key-value store. NAAM's ability to flexibly move message processing by rerouting or retransmitting a message makes it well-suited to solving this temporary interference.
The main challenge in doing this is that NAAM must quickly detect this interference, and its changes to hardware flow steering rules need to take effect quickly to avoid disruption (\S\ref{sec:monitoring}).

Figure~\ref{fig:no-interf} shows the tail latency of a NAAM application running on one host CPU core without interference; Figure~\ref{fig:interf} shows the same application when it must be scheduled with another thread on the same core. This spikes tail latency by three orders of magnitude: from 25~\textmu{}s to 20~ms. In Figure~\ref{fig:interf-shift}, NAAM quickly detects interference for such cases; within 500~ms, it moves execution from the host to the SmartNIC. 

%
%
%
\rs{If we just want to say these two things (about polling, latency currently commented out) we can fit this elsewhere like in the design without dedication a whole subsection heading for it.}

\subsection{The Impact of Placement}

\begin{figure}[t]
\centering
\includegraphics[width=0.8\columnwidth]{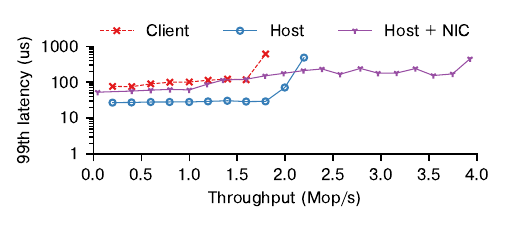}
\caption{Cost of running MICA lookups from different locations.}
\label{fig:placement}
\end{figure}

NAAM messages have the same semantics wherever they run, but the latency and cost of their UDMA operations differ for each application and execution location. Figure~\ref{fig:placement} compares using NAAM messages to perform MICA hash table lookups when messages are run on a client CPU core, on a host CPU core, or on a combination of the host SmartNIC and a host CPU core. Running NAAM messages at the client results in about 3.01 UDMA requests to the host for each NAAM message since it must read at least one hash table bucket along with the correct key-value pair. Running at the host gives better tail response times (2.6-4.0$\times$ speedup) and improves throughput since UDMA operations don't result in multiple round-trips between the client and the host. 
Letting NAAM balance across the SmartNIC and host CPU provides tail response times better than client-side operations (up to 1.3$\times$ speedup). It also allows NAAM to scale beyond what the SmartNIC or host CPU core could each sustain on its own.
\afq{Should we do the graph in linear instead of log scale y?}

\subsection{Handling Faults}

\begin{figure}[t]
\centering
\subfloat[BESS has availability gaps on faults.]{%
    \centering\includegraphics[width=0.8\columnwidth]{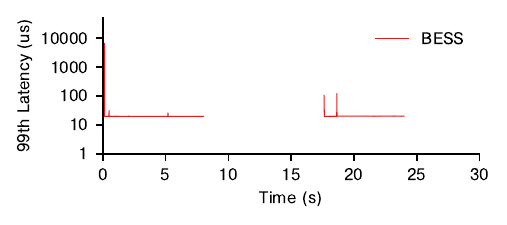}%
    \label{fig:fault-bess}%
}
\vspace{-2eX}
\subfloat[NAAM rejects functions with memory safety bugs.]{%
    \centering\includegraphics[width=0.8\columnwidth]{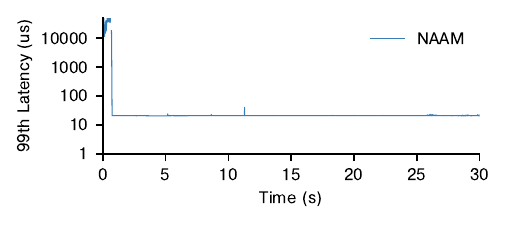}%
    \label{fig:fault-naam}%
}

\caption{NAAM is resistant to faults}
\label{fig:fault-full}
\end{figure}

eBPF's safety guarantees ensure that NAAM functions cannot crash its runtime, which could otherwise cause a denial-of-service by forcing software switch restarts. To demonstrate this, we run an experiment where packets that trigger failures are injected into a forwarding application implemented both as BESS module and as a NAAM application. The function that handles each packet accesses a memory location based on a malformed value found in the packet buffer. This triggers a NULL pointer dereference, which crashes BESS. This function is rejected by NAAM since it does not pass the verifier. Figure~\ref{fig:fault-bess} shows BESS crashing upon receiving the malformed packet and then restarting, which causes it to be unavailable for about 10~s. On the other hand, NAAM is available all the time, even in the face of faulty packets (Figure~\ref{fig:fault-naam}).

\subsection{The Costs of eBPF}

  
\begin{table}[t]
    \footnotesize
    \centering
    \begin{tabular}{l|ccc|ccc}
        \toprule & \multicolumn{6}{c}{\bf Time (ns)} \\
            & \multicolumn{3}{c}{\bf x86-64}
            & \multicolumn{3}{c}{\bf ARMv8} \\
        & & \bf eBPF & \bf eBPF & & \bf eBPF & \bf eBPF \\
        \bf Benchmark &
            \bf Native &
            \bf Interp &
            \bf JIT &
            \bf Native &
            \bf Interp &
            \bf JIT \\ \midrule
         \input{ubench-table-rows} \\ \bottomrule
    \end{tabular}
    \caption{Comparison of Basic NAAM operation latency on x86-64 and ARM in C, in eBPF, and in JITed eBPF code.}
    \label{tbl:ubench}
\end{table}

Finally, to break down the cost that NAAM messages pay to use eBPF for portability and isolation, we profiled several of NAAM's basic operations. Table~\ref{tbl:ubench} shows the results for x86-64 (the client and host architecture) and ARMv8 (the SmartNIC architecture). ``Native'' shows the performance for implementing the listed functionality in C. All UDMA operations are performed against device-local memory (host DRAM on x86-64 and SmartNIC DRAM on ARM).

Our experience and end-to-end benchmarks have suggested to us that each ARM SmartNIC core takes about 5$\times$ longer to perform operations than an x86 core. The results here corroborate that. Both on x86 and ARM eBPF's just-in-time compilation to the target CPU architecture improves the performance of operations substantially from 1.9~to~13.8$\times$.

Finally, the results show that NAAM's fast and JITed scheme for suspending function execution from and to a message's packet buffer is extremely efficient. While function yield and resume accounts for about half of the overhead of a UDMA for an operation that completes against local memory, the cost of suspending and resuming a NAAM function is lower than the cost to set up an RDMA operation (which requires multiple descriptors and DMAs) or the overhead to stall the operation awaiting a response from a remote machine.

\subsection{Cell B-Tree}

\begin{figure}[t]
\centering
\subfloat[B+tree Lookup Latency-Throughput.]{%
    \centering\includegraphics[width=0.8\columnwidth]{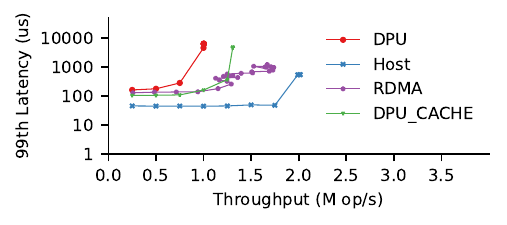}%
    \label{fig:btree-tput-lat}%
}
\vspace{-2eX}
\subfloat[B+tree Lookup Data Transmission Rate.]{%
    \centering\includegraphics[width=0.8\columnwidth]{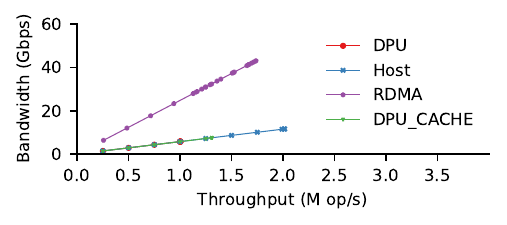}%
    \label{fig:btree-bw}%
}

\caption{B+tree Performance.}
\label{fig:btree-full}
\end{figure}

To demonstrate NAAM's performance benefits in a more complete application, we re-implemented the Cell key-value store's \texttt{GET} operations~\cite{cell}, which execute against an ordered B+tree stored host memory. The original Cell system used a hybrid RDMA and RPC-based access path, and it used response times to decide between using RPCs that traverse the tree or using multiple RDMA read operations that traverse the tree without server CPU involvement. In our reimplementation, we compare the performance of \texttt{GET} operations implemented with RDMA only, using the Host CPU only, and on the DPU only. DPU\_CACHE is a variant that runs \texttt{GET} operations from a cache in DPU memory, which can be done in a consistent fashion by having the DPU flush items which are affected by \texttt{PUT} operations similar to BMC~\cite{ghigoff2021bmc}.

The tree is populated with 10~M eight-byte keys, making the tree five levels deep. Figure~\ref{fig:btree-tput-lat} shows that NAAM messages on the Host CPU sustain 1.75~M GET op/s with \SI{48}{\micro\second} p99 latency. NAAM messages run at the DPU sustain 0.75~M GET op/s with \SI{286}{\micro\second} p99 latency, which improves to 1.3~M GET op/s with \SI{375}{\micro\second} p99 latency when the tree is cached in DPU memory. RDMA sustains 1.3~M GET op/s with \SI{263}{\micro\second} p99 latency. For all of these experiments, the client sends batches of 32~requests (for RDMA these are simply 32 concurrent, pipelined requests). Though RDMA shows similar performance to the DPU when it is using caching, it has an additional cost. RDMA clients must transfer four internal B-tree nodes from the host for each lookup in addition to the leaf page containing the value the client is searching for. Additionally, RDMA-based clients cannot efficiently interpose on all \texttt{PUT} operations issued by all clients for each key. Hence, they cannot efficiently cache the data they read from host memory to avoid this overhead in the same way that DPU\_CACHE does.
Figure~\ref{fig:btree-bw} shows the network bandwidth consumed by each of the approaches. At 1~M op/s, NAAM uses 5.8~Gbps (where operations execute at the server or its SmartNIC) while RDMA uses 25~Gbps ($4.3\times$).

NAAM gives Cell's hybrid \texttt{GET} execution approach more flexibility in where it runs operations. Operations run most efficiently at the Host CPU by avoiding extra round trips and transfers. When the Host CPU becomes congested, NAAM's SmartNIC DPU can use caching for read-intensive workloads which delivers similar performance to the original hybrid Cell approach without wasting network bandwidth. Finally, NAAM could also combine all three approaches by balancing work across Host CPU, DPU, and RDMA.

%% file: ubench-table-rows.tex
Empty Fn & <~1 & 25.8 & 12.4 & <~1 & 103 & 54.7 \\
Fn Yield & - & 91.3 & 14.8 & - & 177 & 54.8 \\
UDMA Rd & 8.7 & 365 & 35.5 & 16.5 & 1511 & 109 \\
UDMA Wr & 11.4 & 399 & 26.7 & 24 & 1536 & 125 

%% file: tex/conclusion.tex
\section{Conclusion}

NAAM is a new approach to network acceleration that lets applications define simple, portable RPC-like functions over remote data. NAAM implements these functions as active messages that are executed in multiple places in the network efficiently. NAAM is efficient since remote memory accesses never stall its runtime, and it can rebalance load using hardware flow steering. Hence, NAAM scales applications across client, server, and SmartNIC in 100s of milliseconds to avoid bottlenecks and route around transient CPU contention.

\section{Acknowledgments}

This material is based upon work supported by the National Science
Foundation under Grant Nos.\ CNS-1750558, CNS-1942686, and CNS-2245999. Any opinions, findings,
and conclusions or recommendations expressed in this material are
those of the authors and do not necessarily reflect the views of the
National Science Foundation. This work was also supported in part by Google LLC
and VMware Research by Broadcom.

%% file: paper.bbl
\begin{thebibliography}{10}

\bibitem{bess}
{BESS: Berkeley Extensible Software Switch}.
\newblock \url{https://github.com/NetSys/bess}.

\bibitem{linux-ebpf}
Linux programmer’s manual. bpf(2).
\newblock \url{https://man7.org/linux/man-pages/man2/bpf.2.html}.

\bibitem{kernel-ebpf}
Linux socket filtering aka berkeley packet filter (bpf).
\newblock \url{https://www.kernel.org/doc/Documentation/networking/filter.txt}.

\bibitem{doca-dma}
Nvidia doca dma programming guide.
\newblock
  \url{https://docs.nvidia.com/doca/sdk/dma-programming-guide/index.html}.

\bibitem{openflow-spec}
{OpenFlow Switch Specification}.
\newblock
  \url{https://opennetworking.org/wp-content/uploads/2014/10/openflow-switch-v1.5.1.pdf}.

\bibitem{rdma-manual}
Rdma aware networks programming user manual v1.7.
\newblock
  \url{https://docs.nvidia.com/networking/display/rdmaawareprogrammingv17}.

\bibitem{ubpf}
ubpf: Userspace ebpf vm.
\newblock \url{https://github.com/iovisor/ubpf}.

\bibitem{AmaroRMC2020}
Emmanuel Amaro, Zhihong Luo, Amy Ousterhout, Arvind Krishnamurthy, Aurojit
  Panda, Sylvia Ratnasamy, and Scott Shenker.
\newblock Remote memory calls.
\newblock In {\em Proceedings of the 19th ACM Workshop on Hot Topics in
  Networks}, HotNets '20, 2020.

\bibitem{asfp}
Ankit Bhardwaj, Chinmay Kulkarni, and Ryan Stutsman.
\newblock {Adaptive Placement for In-memory Storage Functions}.
\newblock In {\em 2020 USENIX Annual Technical Conference (USENIX ATC 20)},
  pages 127--141, 2020.

\bibitem{BrukePrism2021}
Matthew Burke, Sowmya Dharanipragada, Shannon Joyner, Adriana Szekeres, Jacob
  Nelson, Irene Zhang, and Dan R.~K. Ports.
\newblock {PRISM: Rethinking the RDMA Interface for Distributed Systems}.
\newblock In {\em Proceedings of the ACM SIGOPS 28th Symposium on Operating
  Systems Principles}, SOSP '21, 2021.

\bibitem{cavium_intelligent_network_adapters_2023}
Cavium.
\newblock Intelligent network adapters cn63xx nic10e.
\newblock
  \url{http://cavium.com/Intelligent_Network_Adapters_CN63XX_NIC10E.html},
  2023.
\newblock [Online; last accessed 07-Dec-2023].

\bibitem{farm}
Aleksandar Dragojevi{\'c}, Dushyanth Narayanan, Miguel Castro, and Orion
  Hodson.
\newblock {FaRM}: Fast remote memory.
\newblock In {\em 11th USENIX Symposium on Networked Systems Design and
  Implementation (NSDI 14)}, pages 401--414, Seattle, WA, April 2014. USENIX
  Association.

\bibitem{dragojevic2015no}
Aleksandar Dragojevi\'{c}, Dushyanth Narayanan, Edmund~B. Nightingale, Matthew
  Renzelmann, Alex Shamis, Anirudh Badam, and Miguel Castro.
\newblock No compromises: Distributed transactions with consistency,
  availability, and performance.
\newblock In {\em Proceedings of the 25th Symposium on Operating Systems
  Principles}, SOSP '15, page 54–70, New York, NY, USA, 2015. Association for
  Computing Machinery.

\bibitem{du2021fast}
Jingwen Du, Fang Wang, Dan Feng, Weiguang Li, and Fan Li.
\newblock Fast and consistent remote direct access to non-volatile memory.
\newblock In {\em Proceedings of the 50th International Conference on Parallel
  Processing}, ICPP '21, New York, NY, USA, 2021. Association for Computing
  Machinery.

\bibitem{prevail-verifier}
Elazar Gershuni, Nadav Amit, Arie Gurfinkel, Nina Narodytska, Jorge~A Navas,
  Noam Rinetzky, Leonid Ryzhyk, and Mooly Sagiv.
\newblock Simple and precise static analysis of untrusted linux kernel
  extensions.
\newblock In {\em Proceedings of the 40th ACM SIGPLAN Conference on Programming
  Language Design and Implementation}, pages 1069--1084, 2019.

\bibitem{ghigoff2021bmc}
Yoann Ghigoff, Julien Sopena, Kahina Lazri, Antoine Blin, and Gilles Muller.
\newblock $\{$BMC$\}$: Accelerating memcached using safe in-kernel caching and
  pre-stack processing.
\newblock In {\em 18th USENIX Symposium on Networked Systems Design and
  Implementation (NSDI 21)}, pages 487--501, 2021.

\bibitem{gu2017efficient}
Juncheng Gu, Youngmoon Lee, Yiwen Zhang, Mosharaf Chowdhury, and Kang~G. Shin.
\newblock Efficient memory disaggregation with infiniswap.
\newblock In {\em 14th USENIX Symposium on Networked Systems Design and
  Implementation (NSDI 17)}, pages 649--667, Boston, MA, March 2017. USENIX
  Association.

\bibitem{guo2016rdma}
Chuanxiong Guo, Haitao Wu, Zhong Deng, Gaurav Soni, Jianxi Ye, Jitu Padhye, and
  Marina Lipshteyn.
\newblock Rdma over commodity ethernet at scale.
\newblock In {\em Proceedings of the 2016 ACM SIGCOMM Conference}, SIGCOMM '16,
  page 202–215, New York, NY, USA, 2016. Association for Computing Machinery.

\bibitem{kalia2019datacenter}
Anuj Kalia, Michael Kaminsky, and David Andersen.
\newblock Datacenter {RPCs} can be general and fast.
\newblock In {\em 16th USENIX Symposium on Networked Systems Design and
  Implementation (NSDI 19)}, pages 1--16, Boston, MA, February 2019. USENIX
  Association.

\bibitem{kalia2014using}
Anuj Kalia, Michael Kaminsky, and David~G. Andersen.
\newblock Using rdma efficiently for key-value services.
\newblock In {\em Proceedings of the 2014 ACM Conference on SIGCOMM}, SIGCOMM
  '14, page 295–306, New York, NY, USA, 2014. Association for Computing
  Machinery.

\bibitem{design-guildlines}
Anuj Kalia, Michael Kaminsky, and David~G. Andersen.
\newblock Design guidelines for high performance {RDMA} systems.
\newblock In {\em 2016 {USENIX} Annual Technical Conference ({USENIX} {ATC}
  16)}, pages 437--450, Denver, CO, June 2016. {USENIX} Association.

\bibitem{tas}
Antoine Kaufmann, Tim Stamler, Simon Peter, Naveen~Kr Sharma, Arvind
  Krishnamurthy, and Thomas Anderson.
\newblock Tas: Tcp acceleration as an os service.
\newblock In {\em Proceedings of the Fourteenth EuroSys Conference 2019}, 2019.

\bibitem{freeflow}
Daehyeok Kim, Tianlong Yu, Hongqiang~Harry Liu, Yibo Zhu, Jitu Padhye, Shachar
  Raindel, Chuanxiong Guo, Vyas Sekar, and Srinivasan Seshan.
\newblock {FreeFlow}: Software-based virtual {RDMA} networking for
  containerized clouds.
\newblock In {\em 16th USENIX Symposium on Networked Systems Design and
  Implementation (NSDI 19)}, pages 113--126, Boston, MA, February 2019. USENIX
  Association.

\bibitem{splinter}
Chinmay Kulkarni, Sara Moore, Mazhar Naqvi, Tian Zhang, Robert Ricci, and Ryan
  Stutsman.
\newblock {Splinter: Bare-metal Extensions for Multi-tenant Low-latency
  Storage}.
\newblock In {\em 13th USENIX Symposium on Operating Systems Design and
  Implementation (OSDI 18)}, pages 627--643, 2018.

\bibitem{le2019impact}
Yanfang Le, Mojtaba Malekpourshahraki, Brent Stephens, Aditya Akella, and
  Michael~M. Swift.
\newblock On the impact of cluster configuration on roce application design.
\newblock In {\em Proceedings of the 3rd Asia-Pacific Workshop on Networking},
  APNet '19, page 64–70, New York, NY, USA, 2019. Association for Computing
  Machinery.

\bibitem{lim2014mica}
Hyeontaek Lim, Dongsu Han, David~G Andersen, and Michael Kaminsky.
\newblock {MICA: A Holistic Approach to Fast In-Memory Key-Value Storage}.
\newblock In {\em 11th USENIX Symposium on Networked Systems Design and
  Implementation (NSDI 14)}, pages 429--444, 2014.

\bibitem{ipipe}
Ming Liu, Tianyi Cui, Henry Schuh, Arvind Krishnamurthy, Simon Peter, and Karan
  Gupta.
\newblock Offloading distributed applications onto smartnics using ipipe.
\newblock In {\em Proceedings of the ACM Special Interest Group on Data
  Communication}, pages 318--333. 2019.

\bibitem{lu2016high}
Xiaoyi Lu, Dipti Shankar, Shashank Gugnani, and Dhabaleswar~K. Panda.
\newblock High-performance design of apache spark with rdma and its benefits on
  various workloads.
\newblock In {\em 2016 IEEE International Conference on Big Data (Big Data)},
  pages 253--262, 2016.

\bibitem{snap}
Michael Marty, Marc de~Kruijf, Jacob Adriaens, Christopher Alfeld, Sean Bauer,
  Carlo Contavalli, Michael Dalton, Nandita Dukkipati, William~C Evans, Steve
  Gribble, et~al.
\newblock Snap: a microkernel approach to host networking.
\newblock In {\em Proceedings of the 27th ACM Symposium on Operating Systems
  Principles}, 2019.

\bibitem{bpf1993}
Steven McCanne and Van Jacobson.
\newblock {The BSD Packet Filter: A New Architecture for User-level Packet
  Capture.}
\newblock In {\em USENIX winter}, volume~46, pages 259--270, 1993.

\bibitem{openflow}
Nick McKeown, Tom Anderson, Hari Balakrishnan, Guru Parulkar, Larry Peterson,
  Jennifer Rexford, Scott Shenker, and Jonathan Turner.
\newblock Openflow: Enabling innovation in campus networks.
\newblock {\em SIGCOMM Comput. Commun. Rev.}, 38(2):69–74, mar 2008.

\bibitem{mitchell2013using}
Christopher Mitchell, Yifeng Geng, and Jinyang Li.
\newblock Using {One-Sided} {RDMA} reads to build a fast, {CPU-Efficient}
  {Key-Value} store.
\newblock In {\em 2013 USENIX Annual Technical Conference (USENIX ATC 13)},
  pages 103--114, San Jose, CA, June 2013. USENIX Association.

\bibitem{cell}
Christopher Mitchell, Kate Montgomery, Lamont Nelson, Siddhartha Sen, and
  Jinyang Li.
\newblock Balancing {CPU} and network in the cell distributed {B-Tree} store.
\newblock In {\em 2016 USENIX Annual Technical Conference (USENIX ATC 16)},
  pages 451--464, 2016.

\bibitem{mittal2018revisiting}
Radhika Mittal, Alexander Shpiner, Aurojit Panda, Eitan Zahavi, Arvind
  Krishnamurthy, Sylvia Ratnasamy, and Scott Shenker.
\newblock Revisiting network support for rdma.
\newblock In {\em Proceedings of the 2018 Conference of the ACM Special
  Interest Group on Data Communication}, SIGCOMM '18, page 313–326, New York,
  NY, USA, 2018. Association for Computing Machinery.

\bibitem{bertha}
Akshay Narayan, Aurojit Panda, Mohammad Alizadeh, Hari Balakrishnan, Arvind
  Krishnamurthy, and Scott Shenker.
\newblock Bertha: Tunneling through the network api.
\newblock In {\em Proceedings of the 19th ACM Workshop on Hot Topics in
  Networks}, pages 53--59, 2020.

\bibitem{netronome_nfp6xxx_2023}
Netronome.
\newblock Nfp-6xxx product overview.
\newblock \url{https://netronome.com/product/nfp-6xxx/}, 2023.
\newblock [Online; last accessed 07-Dec-2023].

\bibitem{floem}
Phitchaya~Mangpo Phothilimthana, Ming Liu, Antoine Kaufmann, Simon Peter,
  Rastislav Bodik, and Thomas Anderson.
\newblock Floem: A programming system for $\{$NIC-Accelerated$\}$ network
  applications.
\newblock In {\em 13th USENIX Symposium on Operating Systems Design and
  Implementation (OSDI 18)}, pages 663--679, 2018.

\bibitem{poke2015dare}
Marius Poke and Torsten Hoefler.
\newblock Dare: High-performance state machine replication on rdma networks.
\newblock In {\em Proceedings of the 24th International Symposium on
  High-Performance Parallel and Distributed Computing}, HPDC '15, page
  107–118, New York, NY, USA, 2015. Association for Computing Machinery.

\bibitem{nu}
Zhenyuan Ruan, Seo~Jin Park, Marcos~K Aguilera, Adam Belay, and Malte
  Schwarzkopf.
\newblock {Nu: Achieving Microsecond-Scale Resource Fungibility with Logical
  Processes}.
\newblock In {\em 20th USENIX Symposium on Networked Systems Design and
  Implementation (NSDI 23)}, pages 1409--1427, 2023.

\bibitem{aifm}
Zhenyuan Ruan, Malte Schwarzkopf, Marcos~K Aguilera, and Adam Belay.
\newblock {AIFM: High-Performance, Application-Integrated Far Memory}.
\newblock In {\em 14th USENIX Symposium on Operating Systems Design and
  Implementation (OSDI 20)}, pages 315--332, 2020.

\bibitem{flextoe}
Rajath Shashidhara, Tim Stamler, Antoine Kaufmann, and Simon Peter.
\newblock {FlexTOE}: Flexible {TCP} offload with {Fine-Grained} parallelism.
\newblock In {\em 19th USENIX Symposium on Networked Systems Design and
  Implementation (NSDI 22)}, pages 87--102, 2022.

\bibitem{shi2016fast}
Jiaxin Shi, Youyang Yao, Rong Chen, Haibo Chen, and Feifei Li.
\newblock Fast and concurrent {RDF} queries with {RDMA-Based} distributed graph
  exploration.
\newblock In {\em 12th USENIX Symposium on Operating Systems Design and
  Implementation (OSDI 16)}, pages 317--332, Savannah, GA, November 2016.
  USENIX Association.

\bibitem{SidlerStorm2020}
David Sidler, Zeke Wang, Monica Chiosa, Amit Kulkarni, and Gustavo Alonso.
\newblock Strom: Smart remote memory.
\newblock In {\em Proceedings of the Fifteenth European Conference on Computer
  Systems}, EuroSys '20, 2020.

\bibitem{Singhvi1RMA2020}
Arjun Singhvi, Aditya Akella, Dan Gibson, Thomas~F. Wenisch, Monica Wong-Chan,
  Sean Clark, Milo M.~K. Martin, Moray McLaren, Prashant Chandra, Rob Cauble,
  Hassan M.~G. Wassel, Behnam Montazeri, Simon~L. Sabato, Joel Scherpelz, and
  Amin Vahdat.
\newblock 1rma: Re-envisioning remote memory access for multi-tenant
  datacenters.
\newblock In {\em Proceedings of the Annual Conference of the ACM Special
  Interest Group on Data Communication on the Applications, Technologies,
  Architectures, and Protocols for Computer Communication}, SIGCOMM '20, 2020.

\bibitem{mellanox_tile_gx_ug}
Mellanox Technologies.
\newblock Ug130 architectural overview tile-gx.
\newblock
  \url{https://cdn.manesht.ir/17871___210769647-UG130-ArchOverview-TILE-Gx.pdf},
  2012.
\newblock [Online; last accessed 07-Dec-2023].

\bibitem{mellanox_bluefield_smart_nic}
Mellanox Technologies.
\newblock Bluefield smart nic product brief.
\newblock
  \url{http://www.mellanox.com/related-docs/prod_adapter_cards/PB_BlueField_Smart_NIC.pdf},
  2020.
\newblock [Online; last accessed 07-Dec-2023].

\bibitem{tian2021accelerating}
Feng Tian, Yang Zhang, Wei Ye, Cheng Jin, Ziyan Wu, and Zhi-Li Zhang.
\newblock Accelerating distributed deep learning using multi-path rdma in data
  center networks.
\newblock In {\em Proceedings of the ACM SIGCOMM Symposium on SDN Research
  (SOSR)}, SOSR '21, page 88–100, New York, NY, USA, 2021. Association for
  Computing Machinery.

\bibitem{vipin2018fpga}
Kizheppatt Vipin and Suhaib~A. Fahmy.
\newblock Fpga dynamic and partial reconfiguration: A survey of architectures,
  methods, and applications.
\newblock {\em ACM Comput. Surv.}, 51(4), jul 2018.

\bibitem{WANG2023144}
Jing Wang, Chao Li, Yibo Liu, Taolei Wang, Junyi Mei, Lu~Zhang, Pengyu Wang,
  and Minyi Guo.
\newblock Fargraph+: Excavating the parallelism of graph processing workload on
  rdma-based far memory system.
\newblock {\em Journal of Parallel and Distributed Computing}, 177:144--159,
  2023.

\bibitem{wei2018deconstructing}
Xingda Wei, Zhiyuan Dong, Rong Chen, and Haibo Chen.
\newblock Deconstructing {RDMA-enabled} distributed transactions: Hybrid is
  better!
\newblock In {\em 13th USENIX Symposium on Operating Systems Design and
  Implementation (OSDI 18)}, pages 233--251, Carlsbad, CA, October 2018. USENIX
  Association.

\bibitem{xilinx_40_100g_ethernet_2023}
Xilinx.
\newblock 40g/100g ethernet core.
\newblock
  \url{https://www.xilinx.com/products/intellectual-property/40_100g_ethernet.html},
  2023.
\newblock [Online; last accessed 07-Dec-2023].

\bibitem{xilinx_virtex_7_fpga_2023}
Xilinx.
\newblock Virtex 7 fpga family.
\newblock
  \url{https://www.xilinx.com/products/silicon-devices/fpga/virtex-7.html},
  2023.
\newblock [Online; last accessed 07-Dec-2023].

\bibitem{xue2019fast}
Jilong Xue, Youshan Miao, Cheng Chen, Ming Wu, Lintao Zhang, and Lidong Zhou.
\newblock Fast distributed deep learning over rdma.
\newblock In {\em Proceedings of the Fourteenth EuroSys Conference 2019},
  EuroSys '19, New York, NY, USA, 2019. Association for Computing Machinery.

\bibitem{kayak}
Jie You, Jingfeng Wu, Xin Jin, and Mosharaf Chowdhury.
\newblock {Ship Compute or Ship data? Why Not Both?}
\newblock In {\em 18th USENIX Symposium on Networked Systems Design and
  Implementation (NSDI 21)}, pages 633--651, 2021.

\bibitem{zhu2015congestion}
Yibo Zhu, Haggai Eran, Daniel Firestone, Chuanxiong Guo, Marina Lipshteyn,
  Yehonatan Liron, Jitendra Padhye, Shachar Raindel, Mohamad~Haj Yahia, and
  Ming Zhang.
\newblock Congestion control for large-scale rdma deployments.
\newblock In {\em Proceedings of the 2015 ACM Conference on Special Interest
  Group on Data Communication}, SIGCOMM '15, page 523–536, New York, NY, USA,
  2015. Association for Computing Machinery.

\end{thebibliography}
